\documentclass[
    acmtog,
    authorversion=true,
    screen=true,
]{acmart}

\AtBeginDocument{%
	\providecommand\BibTeX{{%
			\normalfont B\kern-0.5em{\scshape i\kern-0.25em b}\kern-0.8em\TeX}}}

\usepackage{lipsum}
\usepackage{graphicx}
\usepackage{subcaption}

\acmSubmissionID{817}

\copyrightyear{2024}
\acmYear{2024}
\setcopyright{rightsretained}
\acmConference[SA Conference Papers '24]{SIGGRAPH Asia 2024 Conference Papers}{December 3--6, 2024}{Tokyo, Japan}
\acmBooktitle{SIGGRAPH Asia 2024 Conference Papers (SA Conference Papers '24), December 3--6, 2024, Tokyo, Japan}\acmDOI{10.1145/3680528.3687657}
\acmISBN{979-8-4007-1131-2/24/12}

\begin{document}
	
        \title{Fa\c{c}AID: A Transformer Model for Neuro-Symbolic Facade Reconstruction}
	
	\author{Aleksander Plocharski}
	\affiliation{%
		\institution{Warsaw University of Technology}
            \country{Poland}} 
          \affiliation{%
		\institution{IDEAS NCBR}            
            \country{Poland}}
            
  	\author{Jan Swidzinski}
        \affiliation{%
		\institution{IDEAS NCBR}            
            \country{Poland}             
            }
  	\author{Joanna Porter-Sobieraj}
	\affiliation{%
		\institution{Warsaw University of Technology}
              \country{Poland}}

	\author{Przemyslaw Musialski}
	\affiliation{%
		\institution{New Jersey Institute of Technology}           
            \country{USA}             
            }
        \affiliation{%
		\institution{IDEAS NCBR}            
            \country{Poland}             
            }

\begin{abstract}
We introduce a neuro-symbolic transformer-based model that converts flat, segmented facade structures into procedural definitions using a custom-designed split grammar. To facilitate this, we first develop a semi-complex split grammar tailored for architectural facades and then generate a dataset comprising of facades alongside their corresponding procedural representations. This dataset is used to train our transformer model to convert segmented, flat facades into the procedural language of our grammar. During inference, the model applies this learned transformation to new facade segmentations, providing a procedural representation that users can adjust to generate varied facade designs. This method not only automates the conversion of static facade images into dynamic, editable procedural formats but also enhances the design flexibility, allowing for easy modifications.
\end{abstract}
	
 

	\begin{teaserfigure}
        \centering
            \includegraphics[width=0.999\textwidth]{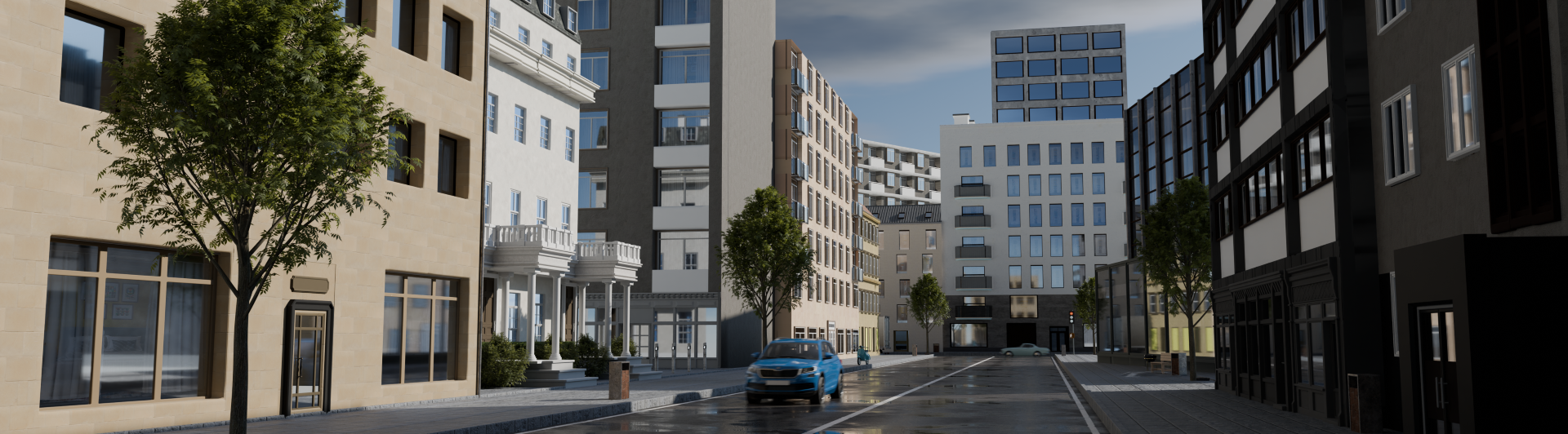}
		\caption{A rendering of a street scene with procedural facades all generated by our method. Novel scene variations can be crafted in the matter of minutes. }
		\label{fig:teaser}
	\end{teaserfigure}
	\maketitle

\begin{figure*}[t]
        \centering
        \includegraphics[trim=0 25 0 0 clip, width=0.98\textwidth]{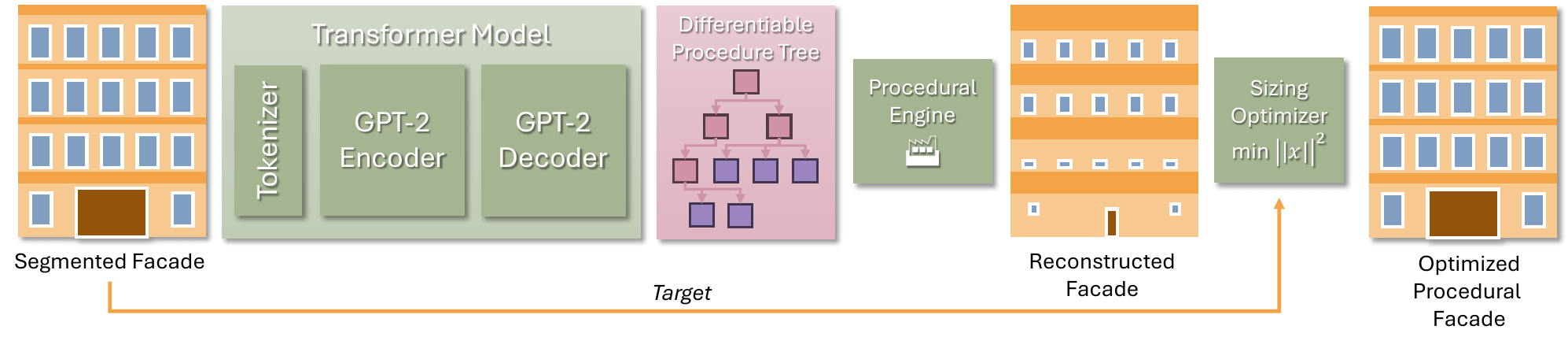}
        \caption{The inference and modeling pipeline of our method: (1) we start with a user provided segmented facade; (2) after preprocessing, a transformer-based model inversely extracts the procedure how to generate the facade; (3) it can be regenerated procedurally, or used for interactive generation of new variations or other facades.}
        \label{fig:pipeline}
\end{figure*}

\newpage
\section{Introduction}\label{sec:intro}
The field of architectural modeling in computer graphics has greatly advanced with the adoption of procedural methods, enabling the efficient generation of vast and complex urban structures and environments~\cite{Wonka2003,Muller2006}. %
However, the creation of grammar rules and their parameters is time-consuming and requires specific knowledge, posing a barrier for non-experts. In response, researchers in computer graphics introduced inverse procedural modeling to extract procedural rules from existing models or images, simplifying the process and making it more accessible~\cite{Musialski2013b,Wu2014,Stava2010}. 

Inverse methods, however, came with new problems, like their complexity, error-proneness, and limited variability and flexibility. To address this, our work introduces a novel neuro-symbolic~\cite{Ritchie2023} approach to enhance facade design. This integration combines neural networks' learning capabilities with procedural models' symbolic reasoning, creating a robust framework for complex architectural modeling. We propose a transformer-based model that converts flat, segmented facade structures into hierarchical procedural definitions using a custom-designed shape grammar~\cite{Stiny1975} tailored for architectural facades~\cite{Wonka2003}. 

Our model is trained on a dataset of segmented facade structures and their corresponding procedural definitions, aiding the model in understanding and replicating structural nuances while aligning with architectural principles. During inference, the model applies learned transformations to new segmented images, converting them into procedural formats. This allows users to generate variations of designs, adjust elements, or create new facades within an established procedural framework, enhancing design flexibility and creativity. Figure~\ref{fig:teaser} shows an urban scene designed by a digital artist using facades generated with our system. 

The key contributions of our work include:   
(1) A neuro-symbolic transformer-based model that automates the conversion of segmented facades into editable procedural definitions. 
(2) A custom-designed split grammar specifically for architectural facades.   
(3) A novel dataset of facade structures and their procedural definitions, serving as a training and validation foundation. 
(4) Demonstrations of the model's practical application in generating varied facade designs, showcasing its utility.

\section{Related Work}\label{sec:related}

\paragraph{Procedural Modeling}
Our work builds on grammar-based procedural methods, introduced for modeling of geometry in botany \cite{Prusinkiewicz1990} and design \cite{Stiny1975}.  %
For modeling of architecture, shape grammars~\cite{Stiny1975} based on splitting rules turned out to be well suitable~\cite{Wonka2003, Muller2006,Whiting2009b}. Such rules decompose geometric shapes into hierarchical structures, making it easier to manage complex  designs. The primary objective of our work is to automatically infer production rules from a grammar given an input segmented facade, which allows for user-friendly editing and generation of new variations~\cite{Lipp2008,Bao2013ProceduralFV,Ilcik2015}. 
\paragraph{Inverse Procedural Modeling}

Inverse procedural modeling aims to derive shape grammars from given segmented geometry or imagery. Early work introduced grammars that split buildings but each rule was crafted and applied manually~\cite{Bekins2005BuildbynumberRT, Aliaga2007}. Subsequent research has continued along this path, proposing various methods to extract procedural descriptions from facades \cite{Muller2007, Becker2009GenerationAA, Teboul2011ShapeGP,Teboul2013ParsingFW,Demir2018GuidedPO}. Notably, the work on inverse procedural modeling of vector-art provided a formal treatment of the problem in graphics \cite{Stava2010}. 
A significant challenge in inverse procedural modeling is dealing with noisy or unsegmented inputs, where lower-level shape understanding and symmetry detection become crucial~\cite{Muller2007, Bokeloh2010}. Once shape grammars are learned from typical input facades, they can be used as priors to guide further reconstruction efforts \cite{Mathias2011Procedural3B, Ripperda2009ApplicationOA, Vanegas2010, Toshev2010DetectingAP, Riemenschneider2012IrregularLF, Teboul2013ParsingFW}.

There have been multiple approaches to extract grammars from simple facades \cite{Weissenberg2013IsTA}, also with varying alternative subdivision rules \cite{Zhang2013LayeredAO}. 
Additionally, the concept of Bayesian models merging to combine deterministic grammars into a stochastic one has been explored, but it necessitates a robust initial grammar \cite{Talton2012LearningDP, Martinovic2013BayesianGL}, or predefined grid-layout structures~\cite{Fan2014StructureCF}. Learning of grammars based on the smallest-grammar paradigm~\cite{Charikar2005TheSG} using approximate dynamic programming~\cite{Wu2013InversePM} has also been proposed. In contrast, our approach relies on pre-defined rules based on a split grammar~\cite{Wonka2003}, but it uses neuro-symbolic transformer model to convert all segmented facade images into procedural representations robustly, streamlining the process of rule applications. 

\paragraph{Generative Models in Graphics}
Generative models have advanced significantly in recent years, particularly in modeling 2D images \cite{Karras2018ASG} and 3D shapes \cite{Nash2020PolyGenAA}. These models often use latent space projections to generate new data \cite{Richardson2020EncodingIS}. 
In the context of building facade images, generative models create realistic urban images, which, although detailed, are often limited in resolution and lack high-level structure or parameterization \cite{Bachl2019CityGANLA, Sun2022AutomaticGO}. Our work addresses these shortcomings by focusing on generating procedural graphs, allowing for enhanced flexibility and customization in facade design.

\paragraph{Neuro-Symbolic Methods in Graphics}
Program synthesis, a concept dating back to 1957 \cite{Backus1957}, explores methods for automatically generating programs from high-level specifications. Inductive synthesis, which relies on partial user specifications and search methods, has seen significant advances in specifying user intent through input-output examples, demonstrations, and natural languages \cite{SolarLezama2006CombinatorialSF, Alur2013SyntaxguidedS}.

In computer graphics, these methods combine the strengths of generative AI and symbolic programs to represent, generate, and manipulate visual and geometric content~\cite{Ritchie2023}. Graph-based models for procedural content generation have been extensively studied, with recursive neural networks capturing dependencies like adjacency and symmetry within shapes \cite{Li2017GRASS, Mo2019StructureNet}. These models have evolved to generate shape programs using recurrent networks and transformers \cite{Jones2020ShapeAssembly, Nash2020PolyGenAA}. In CAD modeling, deep learning has enabled the creation of complex designs from textual descriptions, exemplified by models like DeepCAD \cite{Wu2021DeepCADAD}, and generative models that produce diverse, high-quality procedural materials~\cite{Guerrero2022MatFormer}. Our approach also uses an autoregressive transformer model to convert segmented facade structures into procedural definitions.

Program synthesis transforms modeling into code generation, as seen in reverse-engineering CAD programs from 3D shapes \cite{Du2018InverseCSG, Nandi2018FunctionalPF} and shape program manipulation \cite{Hempel2019SketchnSketchOP}. Recent work in program rewrites generates compact 3D CAD code using E-graphs, optimizing design and fabrication plans \cite{Nandi2019SynthesizingSC}. 
Neuro-symbolic programming combines neural networks with symbolic reasoning to enhance model interpretability and flexibility, applicable across various domains \cite{Chaudhuri2021NeurosymbolicP,Li2023ScallopAL}. 

By integrating neuro-symbolic learning with a custom-designed split grammar, our model automates the conversion of segmented facade structures into procedural definitions, providing a dynamic tool for architects and designers to create and modify facade layouts.

\begin{figure}[t]
        \centering
        \includegraphics[trim=0 0 0 0 clip, width=0.49\textwidth]{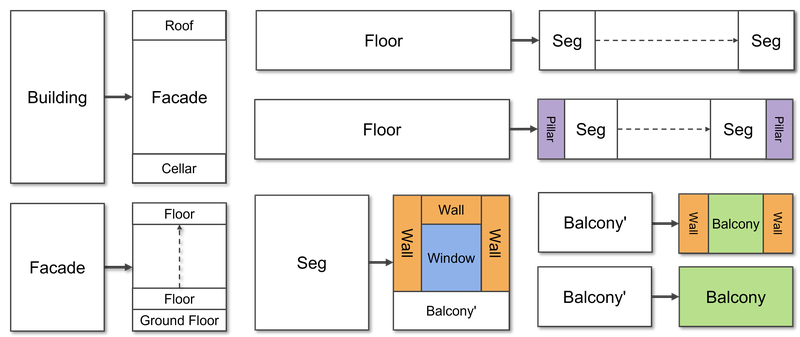}
        \caption{Example production rules of our split grammar which follows the notion as proposed by Wonka et al.~\shortcite{Wonka2003}. The depicted productions are a subset of all productions $P$. For a full set of rules used by our grammar please refer to supplemental material.}
        \label{fig:grammar}        
\end{figure}

\section{Problem Statement and Overview}\label{sec:method}

The main idea of our method is to enable a transformer-based network to infer procedural descriptions from annotated flat grid-like segmentations. The grammar of the procedures is based on split-grammars as proposed by Wonka et al.~\shortcite{Wonka2003} and subsequent work. The control of the grammar is encoded in a heuristic module which we design based on typical architectural rules~\cite{Wonka2003}. 

The main assumption is that an ``oracle'' data generator, composed of a grammar and  heuristics, constitutes the rules. All our data is generated by that generator using stochastic seeding. The resulting procedures are expanded down to the terminal level, which becomes the input to the transformer. In other words, the generator is considered a ``black box,'' whose inverse operation the transformer attempts to learn. 
Thanks to this approach, in contrast to other inverse procedural methods \cite{Wu2013InversePM}, we aim for the procedures to function in a unified space of programs defined by the ``black box,'' rather than requiring the careful construction of a specialized single-use language for each given facade.

The whole system is composed of three main modules (cf. Fig.~\ref{fig:pipeline}):
\begin{enumerate}
    \item A transformer neural model for neuro-symbolic learning, 
    \item a procedural engine, and
    \item an optimization module for sizing adjustments. 
\end{enumerate}

During inference, segmented facade abstractions are passed as input to the transformer model, which predicts a procedure correlating with the underlying structure of the input facade. After this step, a hierarchical procedural representation of all facade elements is generated, but the sizing parameters are not yet set. As the created procedure is differentiable w.r.t. the sizing parameters, a second module—--the optimizer—--minimizes the MSE between the target and the procedural result on a pixel level basis.

\section{Procedural Data Generation}\label{sec:method:data_generation}
To the best of our knowledge, there are no publicly available datasets that contain facades annotated with their procedural definitions of any kind. To create a dataset with facades and their procedural definitions, we developed a split grammar system, drawing on facade modeling techniques from previous work~\cite{Wonka2003}. Please refer to Figure~\ref{fig:grammar} and supplemental material for visual depiction.
A random derivation of the grammar's rule-set will still generate a non-plausible facade in the vast majority of cases. To address this, we implemented a generator that is able to produce a varied distribution of coherent facades based on architectural rules and heuristics. 

The generation process begins with selecting style-specific subsets of architectural parameters. We define them as $S$. They serve as a guideline for keeping the generated design coherent. The generator is then initialized with a starting non-terminal. At each iteration the generator chooses from production rules applicable to the current non-terminal. It employs a heuristic module $H_p(S)$ to select the next proper production to apply. After that another heuristic module, $H_a(S)$, picks the arguments of the selected production rule. The two modules work together to maintain architectural suitability, structural soundness and coherence throughout the whole facade. This sequence of applying rules and adjusting arguments continues until all non-terminals are processed, leaving a complete facade and a derivation tree that documents the creation process, and which can be executed in order to regenerate the facade.

It is those trees of productions that form the language of the procedures we focus on generating. Because of their structure, they are similar to other node-based procedural systems like shader graphs, but instead of performing mathematical calculations, each node partitions the space provided as the input to the node.

The data and the procedural generator are available on the project's website (\href{facaid.github.io}{facaid.github.io}). Please refer to the code for any detail regarding the structure of the grammar as well as the heuristics.

\section{Neuro-Symbolic Generative Model}\label{sec:model}
In this section we describe our core contribution, the transformer-based model for neuro-symbolic generation of procedures from flat facade segmentation. 

\subsection{Transformer-Based Inverse Procedural Learning}\label{sec:model:learning}
Our model is inspired by recent works which utilize transformers~\cite{vaswani2017attention} to represent and generate geometric entities, like plants~\cite{Lee24ToG}, floor-plan layouts~\cite{para2021generative} or apartment layouts~\cite{paschalidou2021atiss,wang2020sceneformer,Leimer2023}. In contrast to the decoder-type models of previous work, our model is comprised of a transformer-based encoder $g_\theta$ which then provides its outputs as context for a transformer-based sequence generator/decoder $h_\phi$. We implement $g_\theta$ and $h_\phi$ as GPT-2 models \cite{radford2019language} using the implementation included in the Hugging Face library \cite{wolf2020transformers}.

\paragraph{Sequence Generation:} A transformer-based encoder $g_\theta$ takes a sequence $S$ and generates a sequence aware embedding $e_i$ for each sequence element $s_i$. Together the embeddings create a new sequence $g_\theta(S) = E$ that encodes the information contained in $S$. 
A transformer-based sequence generator $h_\phi$ factors the probability distribution over sequences $S$ into a product of conditional probabilities over individual tokens:
$$
p(S \mid \phi)=\prod_i p(s_i \mid s_{<i}, \phi),
$$
where $s_{<i}:=s_1, \ldots, s_{i-1}$ denotes the partial sequence up to the token $s_{i-1}$. Given a partial sequence $s_{<i}$, a sequence generator $h_\phi$ predicts the probability distribution over all possible discrete values of the next token: $h_\phi(s_{<i})=p(s_i \mid s_{<i}, \phi)$. The distribution can then be sampled to acquire the value of $s_i$.

Our model additionally conditions the sequence generator $h_\phi$ using the output from the encoder using the cross-attention mechanism extending it to $h_\phi(E, s_{<i})=p(s_i \mid E, s_{<i}, \phi)$ where $E$ is the encoding of the input sequence. The whole model can then be defined as
$$
f_{\theta,\phi}(S_I, s_{<i}) = h_\phi(g_\theta(S_I), s_{<i})
$$
where $S_I$ is the input sequence and $s_{<i}$ are the values from the output sequence that have already been generated.

\paragraph{Tokenization:} In order to feed the data through our model, the inputs and outputs need to be converted into sequences of tokens, respectively $S_I$ and $S_O$. 
The input sequence represents a facade segmentation comprised of non-overlapping rectangles of different classes. Each rectangle $r_i$ is represented by 5 tokens $(t_i, o^x_i, o^y_i, w_i, h_i)$ where $t_i$ indicates the grammar terminal symbol which the rectangle represents, $o^x_i$ and $o^y_i$ are the coordinates of its bottom-left corner and $w_i$ and $h_i$ are respectively its width and height. All of the rectangles $r_i$ are then sorted based on the position of their bottom-left corners (first by the $Y$ coordinate then $X$), in order to decrease the variability of the input sequences that represent the same facades, and concatenated to form the sequence $S_I$.

Formulating the output procedures as sequences $S_O$ poses a greater challenge. Since the procedures take the form of rooted trees we employ the breadth-first traversal method (starting from the root) to produce a sequence of nodes $n_i$. After that we can resume with a similar approach as with the inputs---each node $n_i$ is represented by its class $c_i$ and the set of its structural parameters $(p^1_i, ..., p^j_i)$ which can be integer or floating point values. The nodes $n_i$ are then concatenated with separator tokens in-between them.

When undergoing the tokenization process both categorical and integer parameters are translated into their token counterparts whereas continuous values, like positions and sizes, are discretized between their edge values with a given resolution. After that the model can be trained on all the generated input-output sequence pairs of facade segmentations and their procedural representations.

\paragraph{Positional Encodings:} We employ the usual transformer approach and feed the model with not only the sequences of tokens but also the information about their positions in the sequence. Each token is tagged with its global position in the sequence. Additionally, since our rectangle sequences are represented by groups of 5 tokens we assign each of them a local index from 0 to 4 signifying their location in the sequence of 5~\cite{Leimer2023}. We extend this approach to the output sequences and assign the output tokens a local index in groups that encompass a single production and its arguments. In this case the groups are of dynamic length.

\paragraph{Training:} The tokenized sequence pairs and their positional encodings are then fed through our model in batches. After each batch we perform an optimization step using the AdamW optimizer~\cite{loshchilov2019decoupled}. This process of going through the whole dataset in batches repeats until the loss function reaches a local minimum or the model starts overfitting to the training data.

\begin{figure}[t]
\captionsetup[subfigure]{labelformat=empty}
    \centering
    \hfill
    \begin{subfigure}[t]{0.115\textwidth}
        \includegraphics[width=\textwidth]{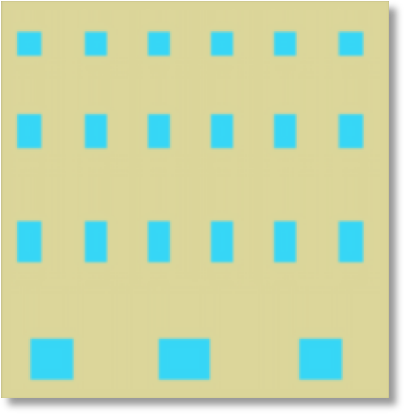}
        \caption{Starting facade}
    \end{subfigure}
    \hfill
    \begin{subfigure}[t]{0.115\textwidth}
    \includegraphics[width=\textwidth]{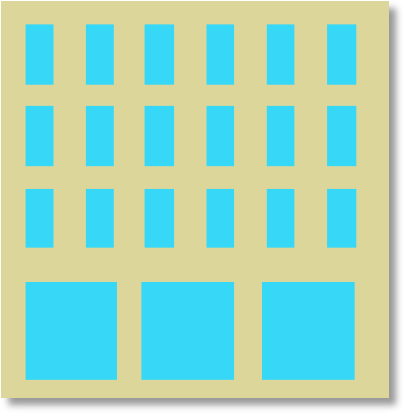}
    \caption{Target facade}
    \end{subfigure}
    \begin{subfigure}[t]{0.115\textwidth}
    \includegraphics[width=\textwidth]{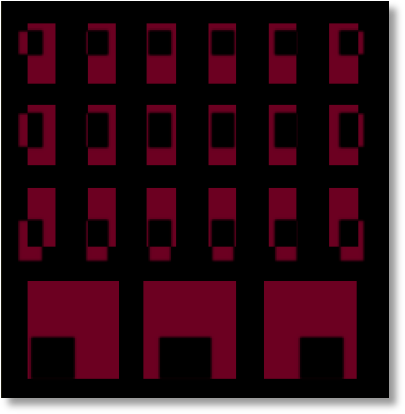}
    \caption{Square error}
    \end{subfigure}
    \begin{subfigure}[t]{0.115\textwidth}
    \includegraphics[width=\textwidth]{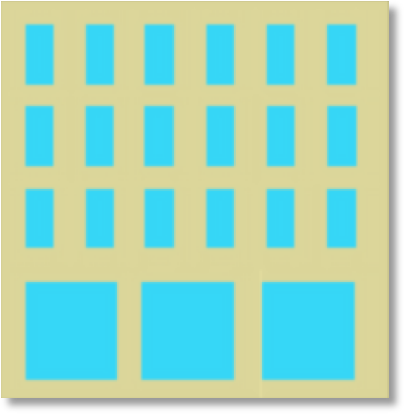}
    \caption{Optimized facade}
    \end{subfigure}
    \hfill
    \caption{Sizing optimizer: The input to sizing optimization is the reconstructed procedure (structural) with default sizing information. As our derived procedures are differentiable w.r.t. the sizing parameters, the terminal sizes are computed by optimizing the square error.}
    \label{fig:optimizer}
\end{figure}

\subsection{Neuro-Symbolic Procedure Inference}\label{sec:model:inference}
The neuro-symbolic generation (inference) is performed by providing a segmented facade image, which can be done using interactive methods~\cite{Musialski2012InteractiveCF} or by using automatic facade segmentation~\cite{Tylecek13} or building labeling methods~\cite{Selvaraju2021BuildingNetLT}. We consider the input facade segmentation is given and represented as a flat irregular 2D grid of boxes with their classes (e.g., \textit{Window}, \textit{Balcony}, etc.) and extents. 

The segmentation is then tokenized using the approach described in Section~\ref{sec:model:learning}. It is crucial that the same tokenization scheme is used here as was used for the training data for the model. The encoding of the input sequence $g_\theta(S_I)$ is calculated and the generation of the output sequence, conditioned on this encoding, begins. The inference is performed one token at a time to make sure the resulting procedure will be syntactically valid and executable. To achieve that at each step of the generation process we perform invalid state nullification---we manually set generated probabilities $p(s_i \mid E, s_{<i}, \phi)$ for syntactically invalid choices to zero and re-normalize the probability distribution. The impact of this step can be seen in Table~\ref{tab:example} which showcases how the average negative log likelihood loss changes after applying this modification for models trained on different amounts of data. The token generation process repeats until an end of sequence token is generated.

The result of the inference is a brand new procedure that re-generates the structure of the flat rectangle representation when executed. Subsequently, the user is free to adjust the parameters of the procedure and create new facade variations.

\begin{figure*}[ht]
    \centering
    \begin{subfigure}[b]{0.59\textwidth}
        \includegraphics[width=\textwidth]{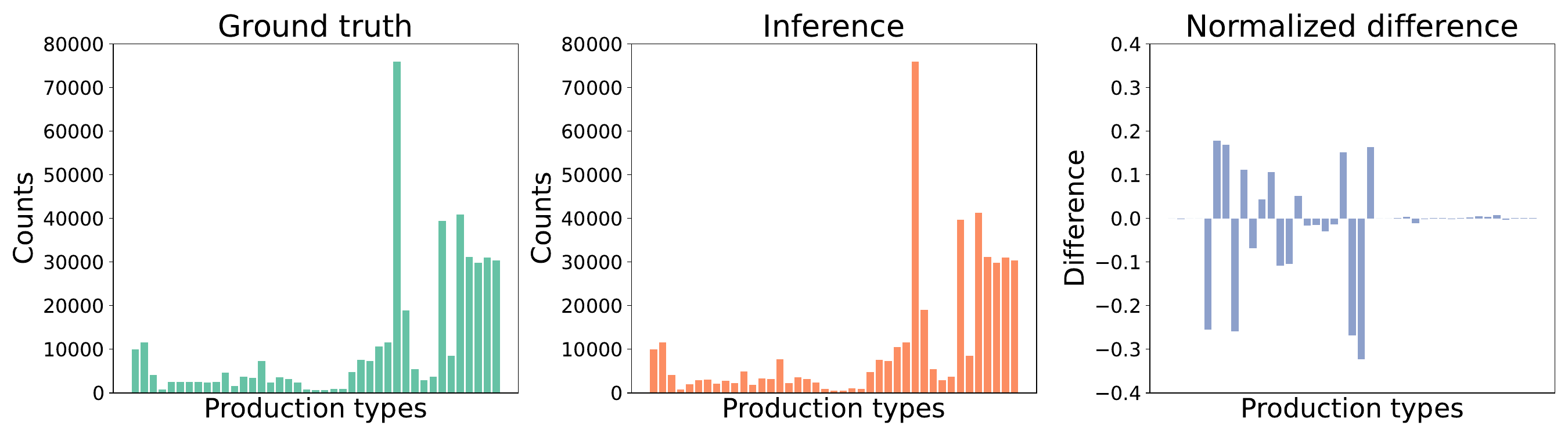}
        \caption{Production Reconstruction Frequency }
        \label{fig:frequency}
    \end{subfigure}
    \hfill
    \begin{subfigure}[b]{0.39\textwidth}
    \centering
        \includegraphics[height=2.95cm]{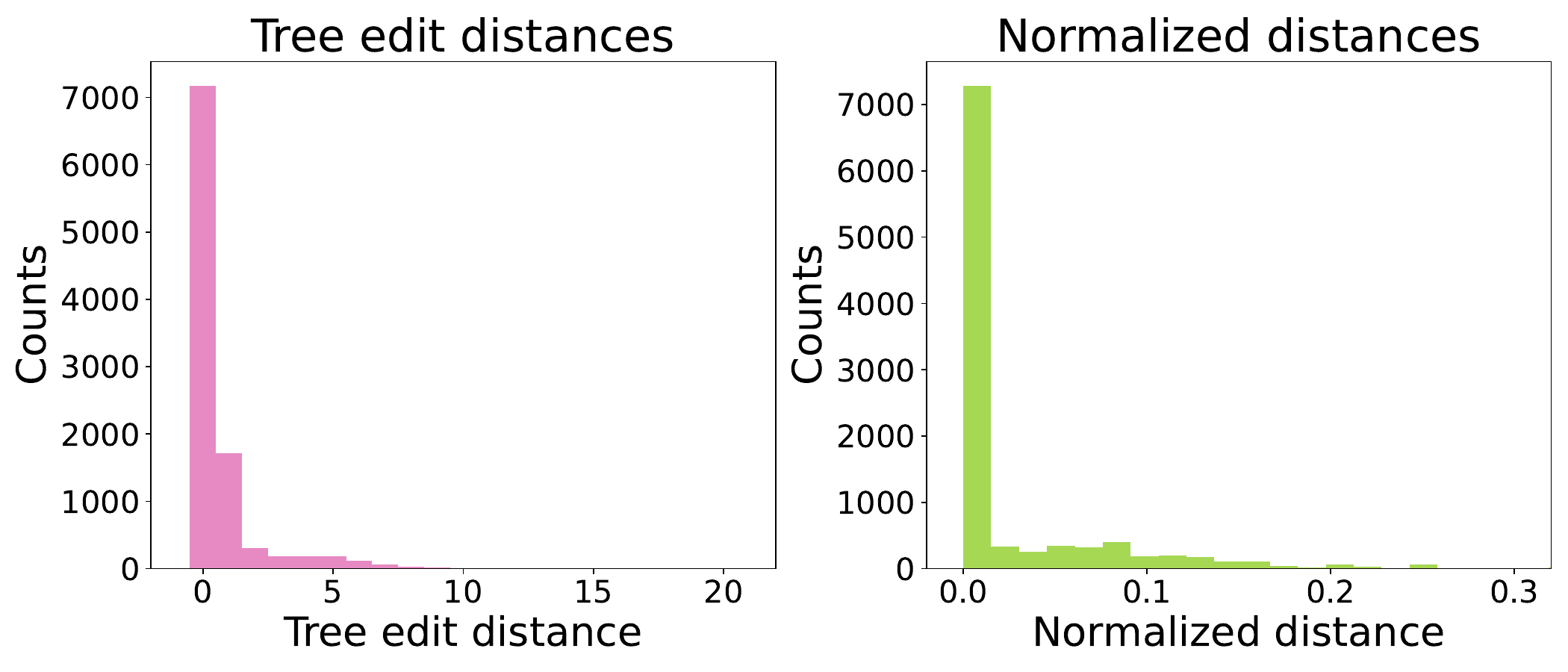}
        \caption{Structural Difference (Tree Edit Distance)}
        \label{fig:edit_dist}
    \end{subfigure}
    \caption{Quantitative evaluation of our model: The first metric shows the frequency of the usages of the procedural rules between the ground truth and reconstructions as well us their difference normalized by inference counts. The second metric measures how much tree-edits are needed to change the reconstructed rules to those from ground truth. The normalized distances represents edit distances divided by their original tree size.}
    \label{fig:overall}
\end{figure*}

\begin{figure}[t]
        \centering
        \begin{subfigure}[t]{0.22\textwidth}
            \includegraphics[trim=0 30 0 0 clip, width={\textwidth}]{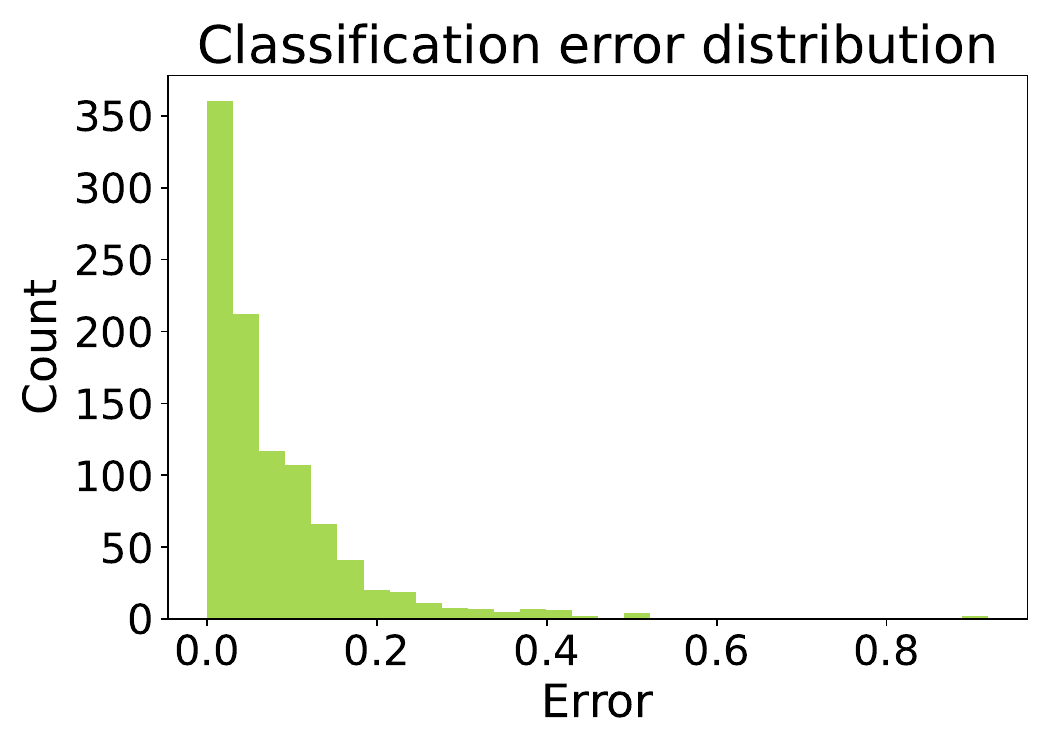}
            \caption{Optimized facades classification error distribution. The error compares the input and output segmentations (pixelwise).}
            \label{fig:optim_error}
        \end{subfigure}
        \hfill
        \begin{subfigure}[t]{0.22\textwidth}
            \includegraphics[trim=0 30 0 0 clip, width={\textwidth}]{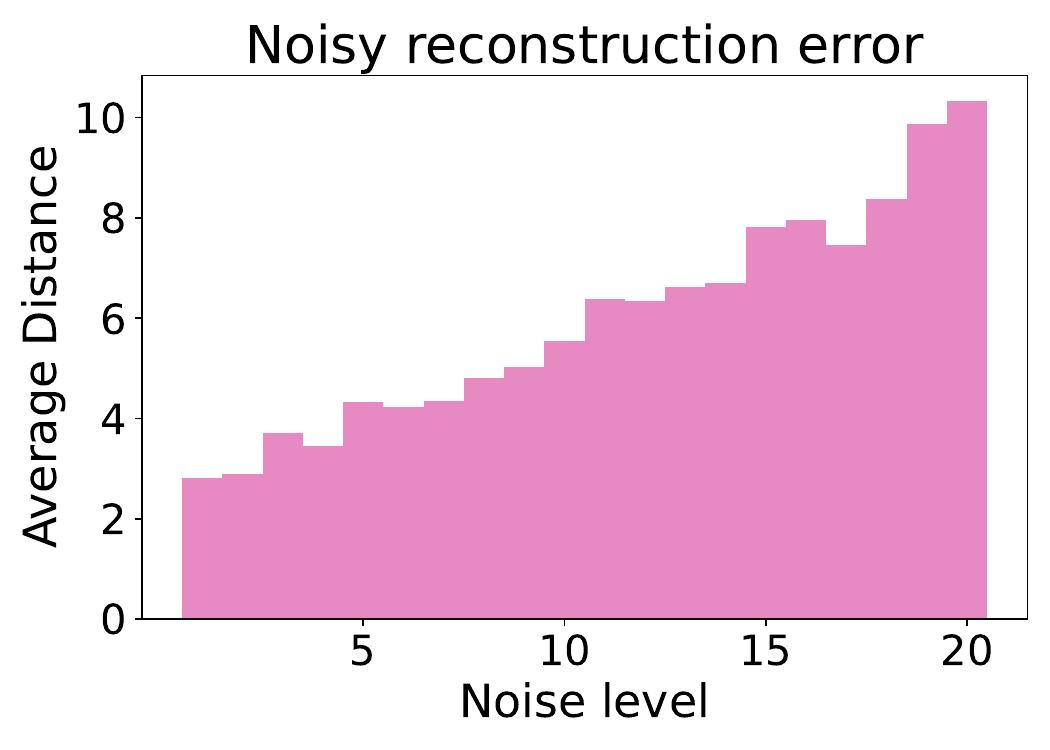}
            \caption{Average tree edit distance to the ground truth when performing inference using inputs with a given level of noise.}
            \label{fig:noise}
        \end{subfigure}
        \caption{Facade reconstruction error}
\end{figure}

\begin{table}[b]
    \centering
    \caption{The average NLL loss of our model on a test set of 10k facades. Both rows show the test loss obtained using models trained on different amounts of data (50k, 100k and 200k respectively) with and without Invalid State Nullification (refer to Section~\ref{sec:model:inference}).}
    
    \begin{tabular}{|c|c|c|c|}
        \hline
        & \textbf{50k} & \textbf{100k} & \textbf{200k} \\
        \hline
        \textbf{without nullification} & 0.61404 & 0.34671 & 0.21412 \\
        \hline
        \textbf{with nullification} & 0.06301 & 0.04868 & 0.02519 \\
        \hline
    \end{tabular}
    \label{tab:example}
\end{table}

\subsection{Differentiable Procedure Sizing Optimization}

The model inference process leaves us with a structurally sound representation of the input facade but all of its sizing parameters are set to default values. In order to fully reproduce the input facade the procedure needs to go through the optimization module which finds the proper values of the sizing parameters. This module heavily leverages the fact that we have implemented all of our production rules in a way that makes them differentiable w.r.t. their sizing parameters. This means that when the procedure is executed all positions and sizes of resulting rectangles can be expressed as differentiable functions of the parameters of the procedure. We use this property to perform a gradient based optimization.

In order to perform the optimization we define a loss function that enables the model to sequentially approach the input facade segmentation. We treat the input segmentation as an image and then calculate the mean square error between it and the result of executing our procedure (Figure~\ref{fig:optimizer}). The engine result also needs to be rasterized since it's a sequence of rectangles. To keep the function differentiable we rasterize all of the rectangles in a soft manner---using a combination of sigmoid functions at the edges of the rectangles. This allows us to calculate the gradient of the mean square error w.r.t. all of the sizing parameters of the procedure.

We use the Adam optimizer\cite{kingma2017adam} to iteratively minimize the loss and optimize the sizing parameters. The process ends once the loss function reaches a local minimum. After that we are left with a procedure that when executed matches the input flat rectangle representation exactly, while also further allowing the user to generate similar variations of the design. 

\begin{figure*}[t]
    
    \centering
    \begin{subfigure}[t]{0.24\textwidth}
        \includegraphics[width=\textwidth]{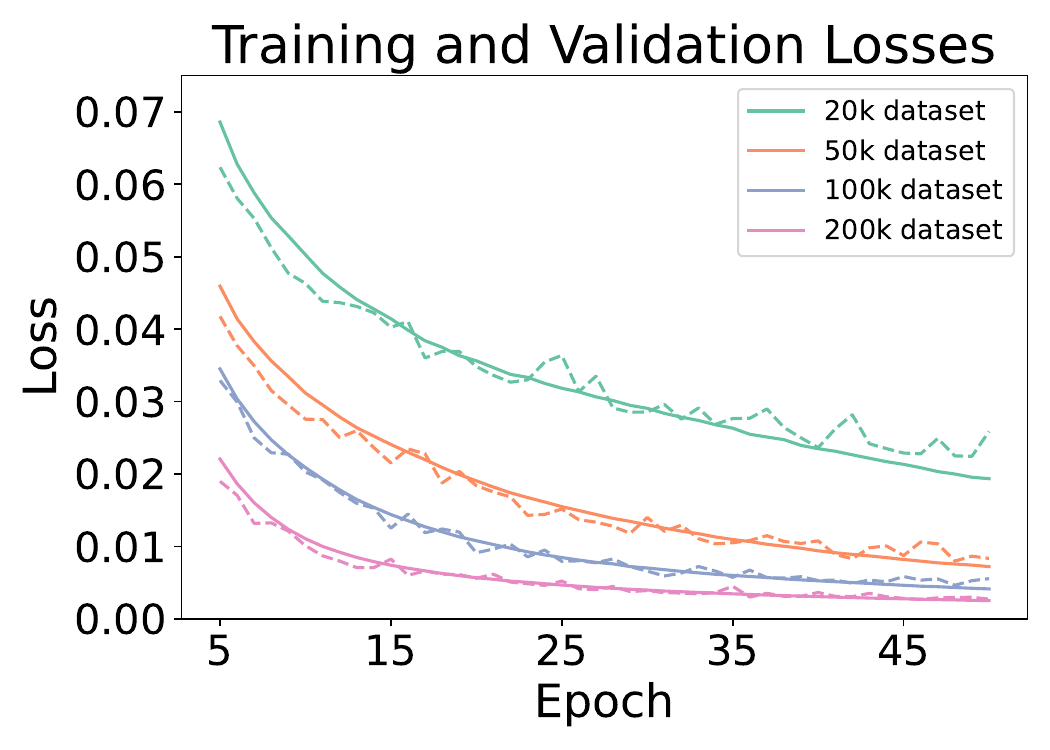}
        \caption{Size of the training set and its influence on the loss. All training data sampled from the same distribution.  }
        \label{fig:subfig1:train_size}
    \end{subfigure}
    \hfill
    \begin{subfigure}[t]{0.24\textwidth}
        \includegraphics[width=\textwidth]{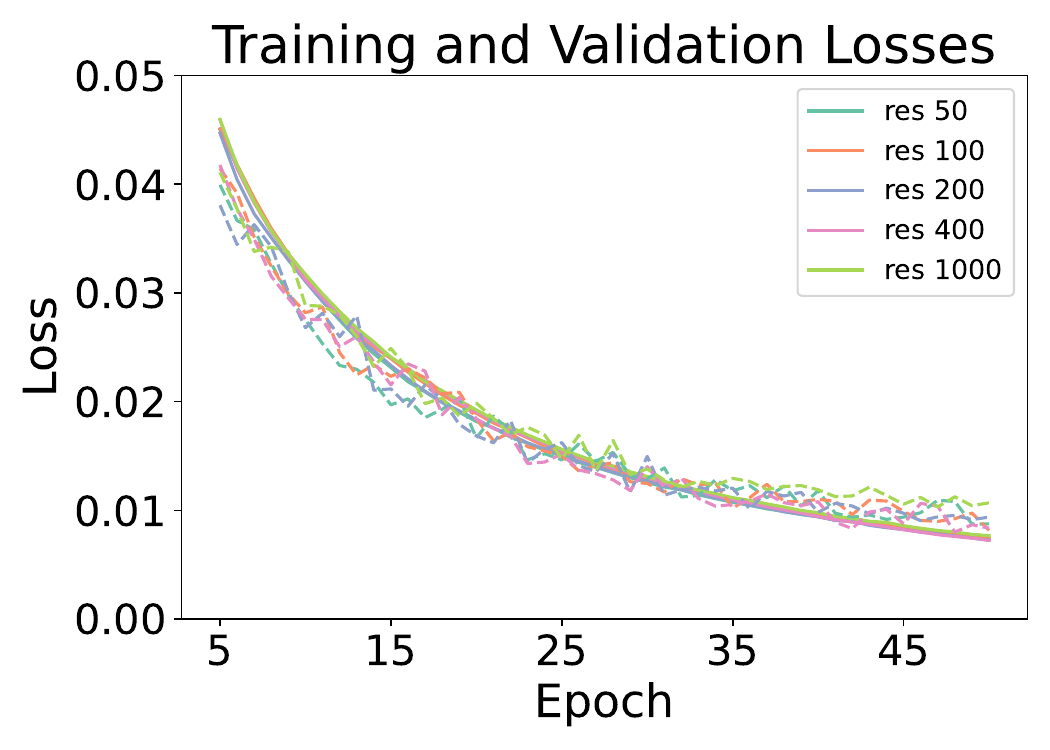}
        \caption{Resolution of the tokenization. Our model is mostly agnostic w.r.t. to that.  }
        \label{fig:subfig2:tokenization}
    \end{subfigure}
    \hfill
    \begin{subfigure}[t]{0.24\textwidth}
        \includegraphics[width=\textwidth]{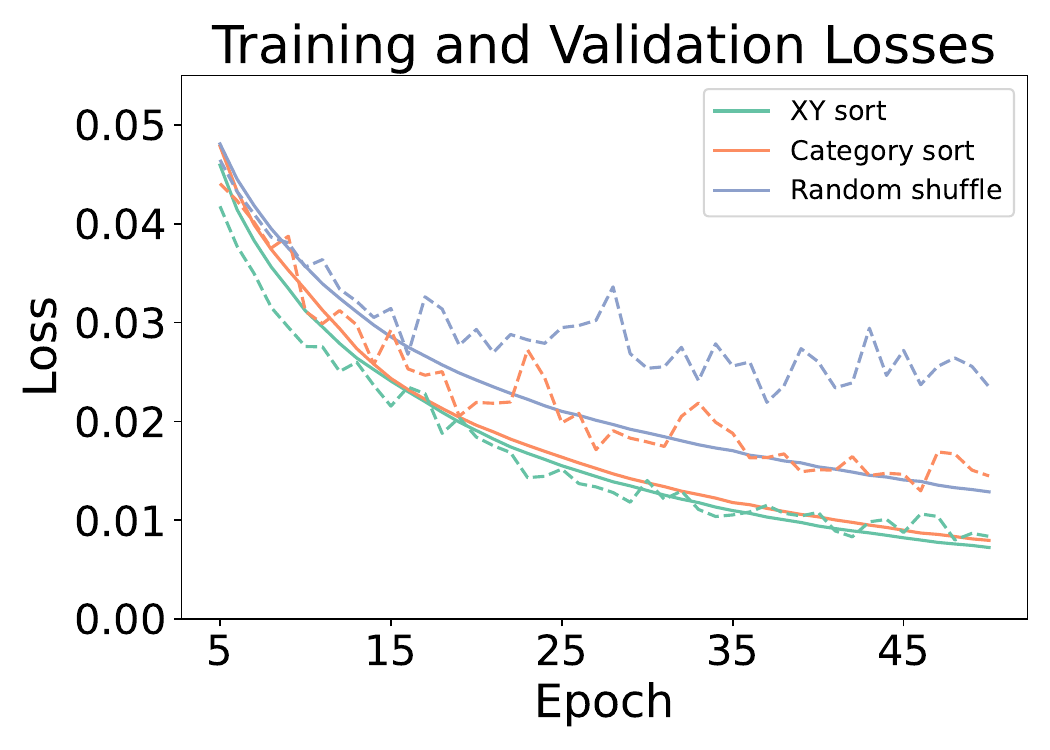}
        \caption{Sorting of the input shapes after segmentation.} %
        \label{fig:subfig3:sorting}
    \end{subfigure}
    \hfill
    \begin{subfigure}[t]{0.24\textwidth}
        \includegraphics[width=\textwidth]{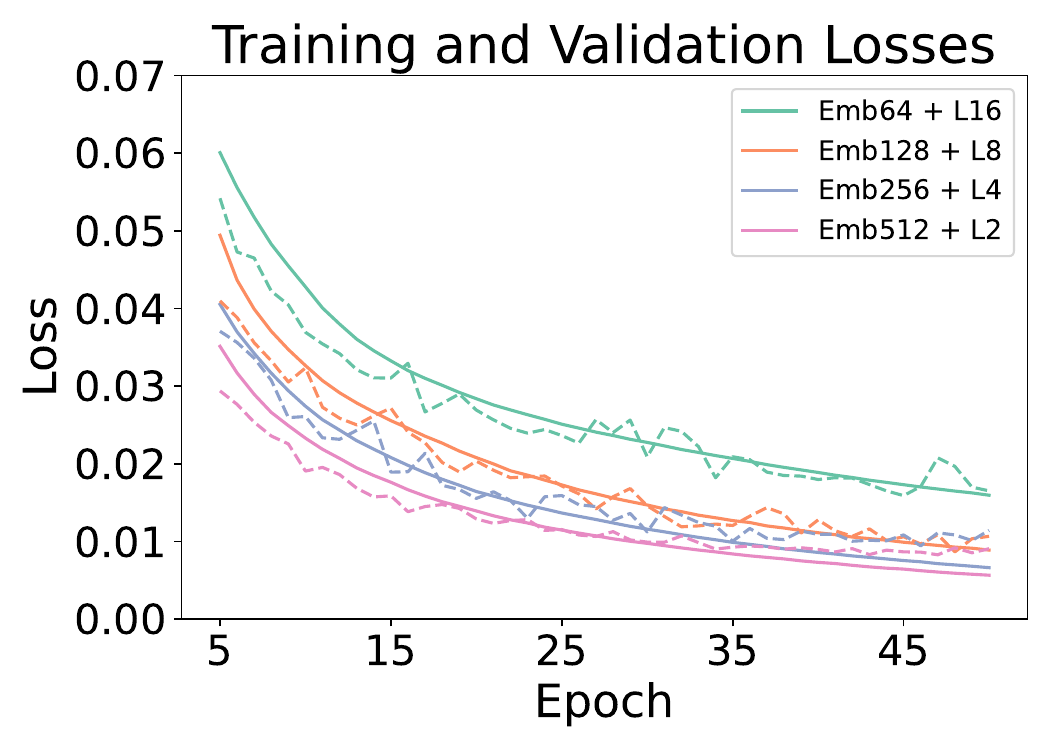}
        \caption{Influence of the model's embedding dimensions (Emb\#) and number of layers (L\#). } %
        \label{fig:subfig4:depth_width}
    \end{subfigure}

    \begin{subfigure}[t]{0.24\textwidth}
        \includegraphics[width=\textwidth]{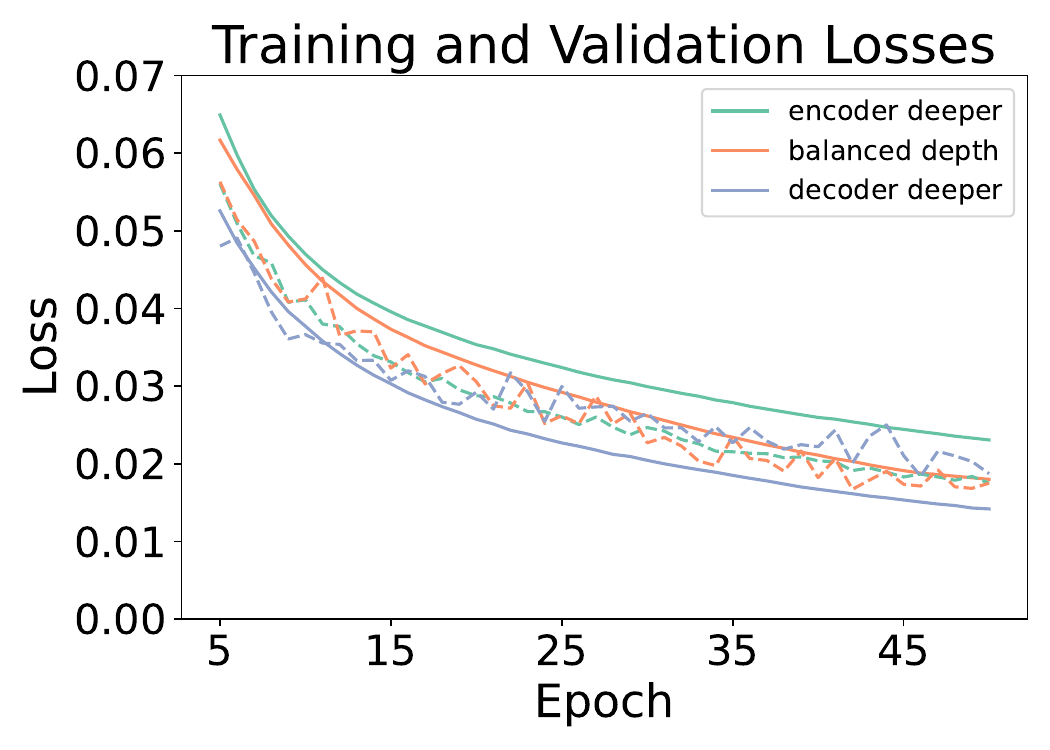}
        \caption{Encoder vs decoder depth.  }
        \label{fig:subfig5:enc_dec_depth}
    \end{subfigure}
    \hfill
    \begin{subfigure}[t]{0.24\textwidth}
        \includegraphics[width=\textwidth]{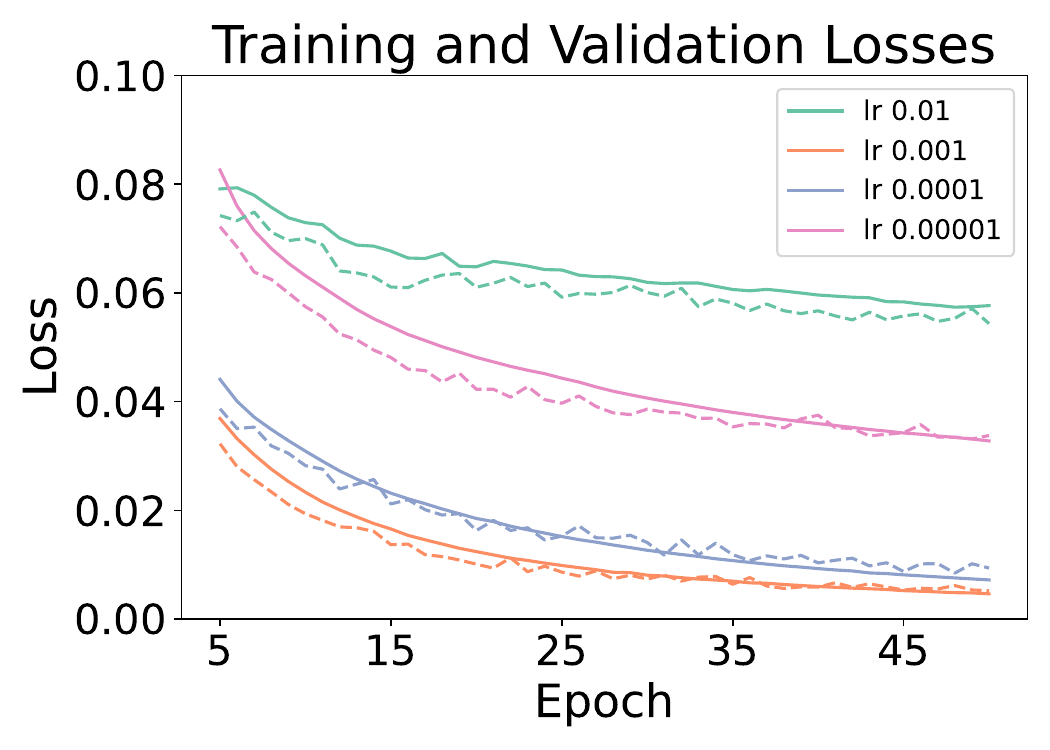}
        \caption{Influence of the learning rate. }
        \label{fig:subfig6:lr}
    \end{subfigure}
    \hfill
    \begin{subfigure}[t]{0.24\textwidth}
        \includegraphics[width=\textwidth]{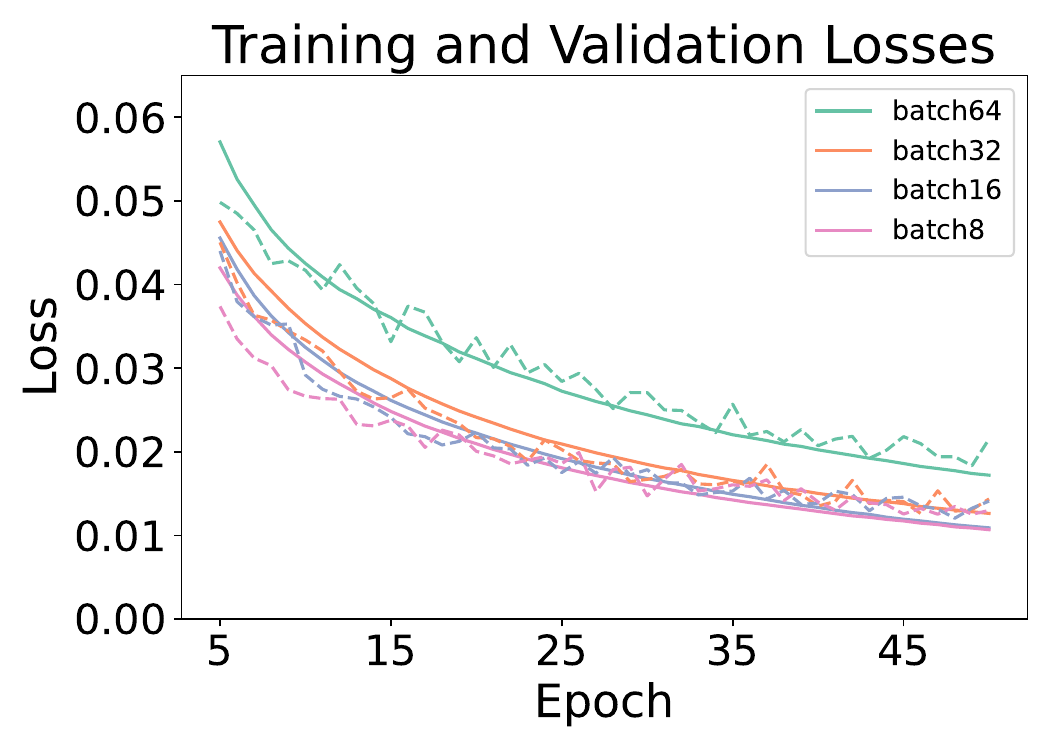}
        \caption{Influence of the batch size.}
        \label{fig:subfig7:batch}
    \end{subfigure}
    \hfill
    \begin{subfigure}[t]{0.24\textwidth}
        \includegraphics[width=\textwidth]{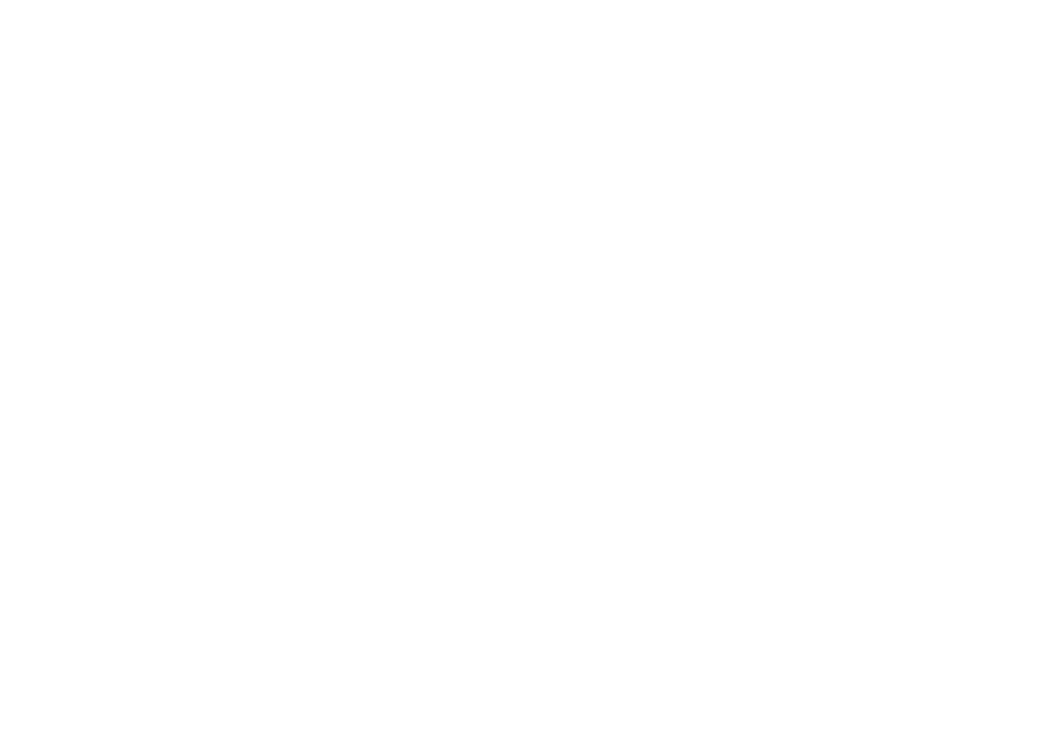}
    \end{subfigure}
    
    \caption{Ablations of our model during training and validation measured using the teacher forcing cross-entropy loss. The solid curves show the training loss and the dashed curves show the validation loss. For a more detailed discussion of the ablation refer to Section~\ref{sec:ablations}. Please note, plots start at epoch 5. }
    \label{fig:ablations1}
\end{figure*}

\subsection{Implementation and Training}
We have implemented our system using Python and the PyTorch library. 
We trained our model using datasets of varying sizes (50k, 100k, 200k) on a NVIDIA L4 24GB and the NVIDIA A100 40GB graphics cards. The training time for 50k samples was 13h on the L4, and 2.8 times faster on the A100 (5h, 10h, 20h respectively for each  training data size). 
For most of our subsequent quantitative results we used the 50k dataset, as it has the same convergence behavior as the bigger sets (cf. Fig~\ref{fig:subfig1:train_size}), however, the qualitative results presented below were generated using the 200k samples dataset.
Additional material is available on \href{facaid.github.io}{facaid.github.io}.

\section{Quantitative Evaluation}\label{sec:quantitative_eval}

In this section we evaluate the performance of the model's reconstruction abilities of the ground truth grammar. It is a key evaluation which shows that a transformer model can reconstruct a semi-complex split grammar, as proposed in Section~\ref{sec:method:data_generation}.

\paragraph{Procedure Reconstruction Frequency}

To perform the structural evaluation we have reconstructed 10k facade structures from a test dataset, never before seen by our model, generated using the data generator described in Section~\ref{sec:method:data_generation}. In Figure~\ref{fig:frequency} we present the usage frequency of productions of our split-grammar in the 10k facades dataset. First, from the ground truth derivation trees and second, from the trees generated by our model. We observe that the model is able to closely approximate the behavior and heuristics of the data generator. The differences highlighted on the third plot mostly arise from the model alternating between productions that have similar visual effect which suggests its generalization abilities when reconstructing geometrical structures.

\paragraph{Procedure Structural Difference: Tree Edit Distance}

The structural reconstructions are also measured against the ground truth based on the number of edits required to arrive at a precise reconstruction of the ground truth derivation tree. We use tree edit distance as our metric of the edits assuming that a single edit can amount to the following changes in the tree: (1) removing a node, (2) adding a node, (3) relabeling a node (changing its production type). Figure~\ref{fig:edit_dist} shows that the majority of our structural reconstructions does not require any edits, while a vast majority requires no more than 5 editing steps. This, in most cases, means that less than 10\% of the total tree needs to be modified to arrive at the ground truth structure.

\paragraph{Reconstruction Classification Error}

We perform the quantitative measure of the optimized module on 1k, of the generated 10k, facade structures. The structures are optimized against a target segmentation from the ground truth. The metric used to compare the result with the input facade calculated what percentage of the segmentation does not match the ground truth exactly. In most cases the optimizer is able to adjust the parameters of the structural representations almost exactly, while a vast majority of the facades has an at least 80\% match with the ground truth (cf. Fig.~\ref{fig:optim_error}).

\paragraph{Robustness to Noise in the Input}

We have tested the robustness of our model when provided with noisy segmentations. We generate a test set of 500 procedural facades, never seen before by the model. After that we apply different levels of noise to all elements of each segmentation. The level of noise is calculated based on the percentage of pixels in the new segmentation which differ from the original one. We check the quality of the structural reconstruction of our model by comparing the inference results generated from the noisy inputs with the ground truth procedures using tree edit distance. Figure \ref{fig:noise} showcases the average tree edit distance for a given noise level. The model is able to reconstruct segmentations with up to 10\% noise with a good level of accuracy---less than 6 edits to arrive at the ground truth.

\section{Qualitative Results}

\subsection{Comparison to Previous Methods}
\begin{figure}[b]
    \centering
    
    \begin{subfigure}[t]{0.45\textwidth}
        \begin{minipage}[b]{3cm}
            \centering
            \textbf{Input}\\
            \vspace{1.4mm}
            \includegraphics[height=2cm]{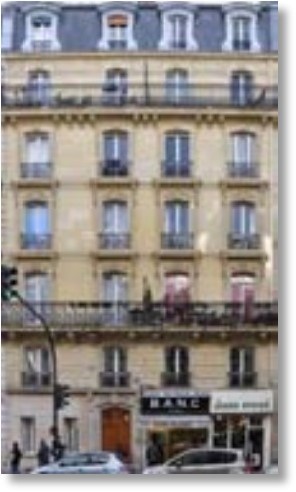}
            \hspace{3mm}
            \includegraphics[height=2cm]{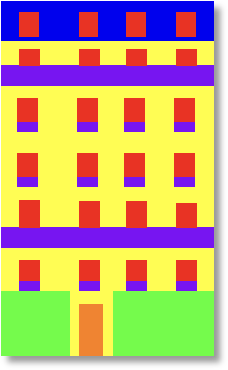}
        \end{minipage}
        \hfill
        \rule{0.4pt}{2cm}
        \hfill
        \begin{minipage}[b]{1.2cm}
        \centering
        \textbf{Theirs}\\
        \vspace{2mm}
        \includegraphics[height=2cm]{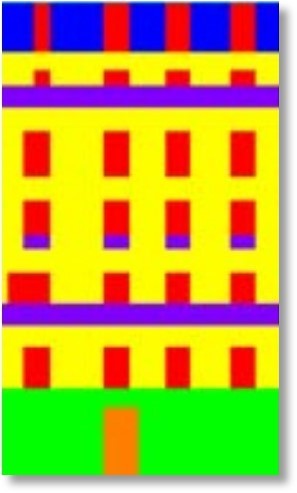}
        \end{minipage}
        \hfill
        \rule{0.4pt}{2cm}
        \hfill
        \begin{minipage}[b]{2cm}
        \centering
        \textbf{Ours}\\
        \vspace{2mm}
        \includegraphics[height=2cm]{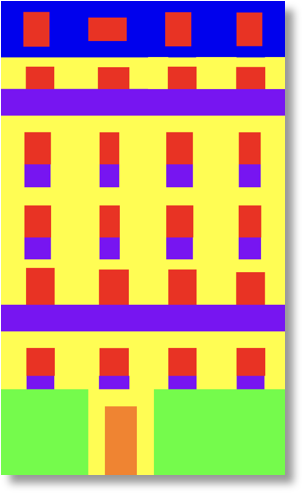}
        \end{minipage}
        \caption{\cite{Riemenschneider2012a}}
    \end{subfigure}
    
    \begin{subfigure}[t]{0.45\textwidth}
        \vspace{3mm}
        \begin{minipage}[b]{3cm}
            \centering
            \includegraphics[height=2cm]{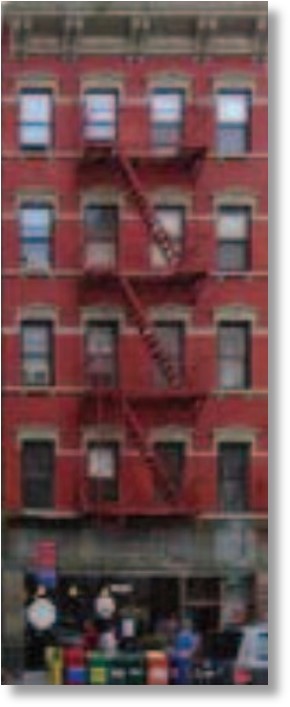}
            \hspace{1.1mm}
            \includegraphics[height=2cm]{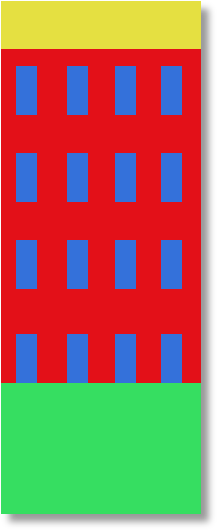}
            \hspace{1.1mm}
            \includegraphics[height=2cm]{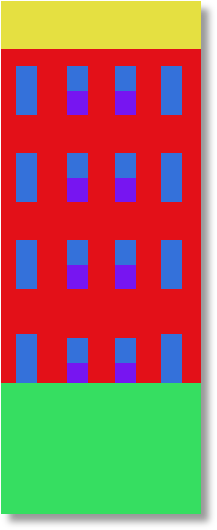}
        \end{minipage}
        \hfill
        \vline
        \hfill
        \begin{minipage}[b]{1.2cm}
            \centering
            \includegraphics[height=2cm]{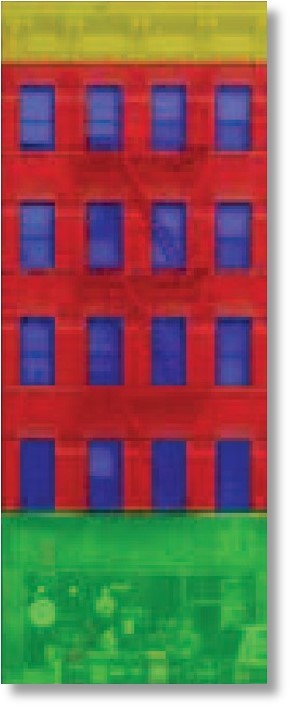}
        \end{minipage}
        \hfill
        \vline
        \hfill
        \begin{minipage}[b]{2cm}
            \centering
            \includegraphics[height=2cm]{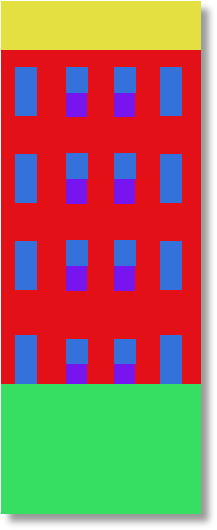}
            \hspace{1.5mm}
            \includegraphics[height=2cm]{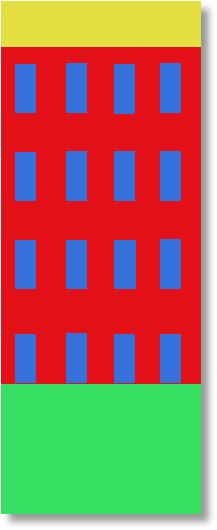}
        \end{minipage}
        \caption{\cite{Teboul2013ParsingFW}}
        \label{fig:comp_teboul}
    \end{subfigure}
    
    \caption{Comparison to the work of Riemenschneider et~al.~\shortcite{Riemenschneider2012a} and Teboul et~al.~\shortcite{Teboul2013ParsingFW}. The first column shows the original facade image and its segmentation(s). The second column displays the results of their methods. The last column shows the segmented facades produced by our method.}
    \label{fig:qual_comp}
\end{figure}

We compare the results of our method to the work of Riemenschneider et~al.~\shortcite{Riemenschneider2012a} and Teboul et~al.~\shortcite{Teboul2013ParsingFW} by using the same target facades and segmenting them by hand in a similar way as as shown in their papers. Figure~\ref{fig:qual_comp} showcases the respective results. Our method demonstrates the capability to reproduce procedural representations which are comparable to their results. Note that the results are highly dependent on the input segmentations, the creation of which can be ambiguous: if we include the balconies explicitly in the segmentation they are recognized by our algorithm; if not, they are missing in our result, producing the same output as their method (Figure~\ref{fig:comp_teboul}). Note that their results were taken directly from the papers because of the lack of available implementations.

\subsection{Qualitative Evaluation}

The qualitative results were generated using manually segmented and annotated facades. The segmentations have been crafted such that they match the terminals of our grammar and take around 5 minutes each to construct. For that task we have implemented a rudimentary segmentation interface. In practice, our goal would be to either use automatic segmentation tools, based on machine learning methods~\cite{Selvaraju2021BuildingNetLT} or on interactive segmentation tools~\cite{Musialski2012InteractiveCF}. However, generally, we consider that as given input which is not part of and essentially does not influence the contribution of this paper. 

\begin{figure*}[p]
        \centering
        \includegraphics[trim=0 500 0 100 clip, width=0.99\textwidth]{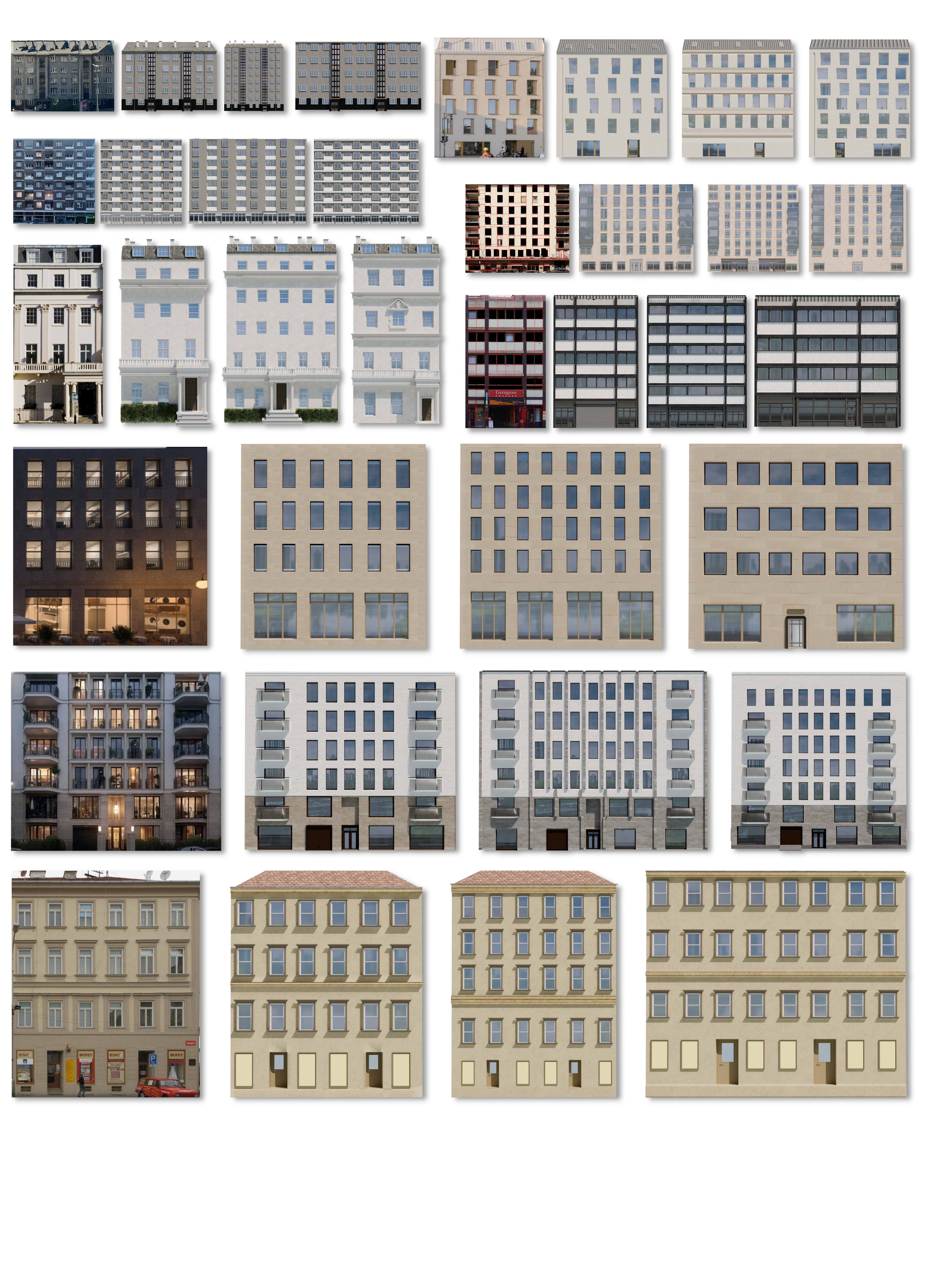}
        \caption{Rendering of the results of our method including material properties assigned to different segmentation classes. Each sequence shows the input image, the procedural reconstruction, and two variations of the same facade procedure. Best viewed in close-up in the electronic version.  }
        \label{fig:rendered_results}
\end{figure*}

The results of the inference of the segmented inputs are depicted in Figure~\ref{fig:qualitative_results}. 
Each facade is comprised of 5 images: (1) image of the facade being reconstructed, (2) the structural representation generated by the transformer model with default sizing parameters, (3) an optimized facade which sizing now matches the target facade, (4) and (5) variations of the target facade created by slightly modifying the procedural representation of it. 
The results show that our approach can successfully reproduce facades of different structures and styles while giving us the ability to generate brand new variations with little to no effort. Additionally, Figure~\ref{fig:rendered_results} showcases renderings of the facades based on the resulting segmentations.

\section{Model Ablations}\label{sec:ablations}
In this section we evaluate the performance of our model depending on various ablations of the setup; the charts are depicted in Fig.~\ref{fig:ablations1}. 

\paragraph{Tokenization Resolution}

We have trained our model on 5 different segmentation discretization resolutions. Starting from a fine grid---1000x1000, all the way to a very coarse one---50x50. The results (cf. Figure~\ref{fig:subfig2:tokenization}) show that the resolution has little to no impact on the model's ability to learn. We assume that this might be caused by the fact that the model learns the structure mostly based on the order and counts of the rectangles that form the input segmentation and their geometric parameters (positions and sizes) just offer additional guidance in which case even the coarse discretization grid should be sufficient. After all, for a single facade structure, there exist multiple sizing variations.

\paragraph{Encoder Depth vs Decoder Depth}

Since the model is composed of two distinct elements---encoder and decoder---we have tested whether making one of them bigger than the other would be beneficial to the learning process. We observe that the model with a deeper decoder tends to overfit a bit faster while making the encoder deeper produces the opposite result (cf. Figure~\ref{fig:subfig5:enc_dec_depth}). That is why we opt for a balanced architecture, as the validation curve of such a model correlates most closely with its training counterpart. 

\paragraph{Embedding Dimensions vs Number of Layers}

The model was also tested against its overall width and depth. We have varied those parameters by a factor of 2 for each of them. The results show that the model performs best when the number of its embedding dimensions (width) greatly surpasses its number of layers (cf. Figure~\ref{fig:subfig4:depth_width}). We conclude that this result most likely comes from the fact that the transformer model architecture already provides a lot of depth by feeding the embeddings through the quite shallow blocks multiple times. This allows the embeddings to pick up additional context information while their larger size provides the space required to store more of the nuances of the segmentation structure.

\paragraph{Input Segmentation Rectangle Sorting}

As mentioned in Section~\ref{sec:model}, to provide the model with a structured format of the input segmentations we sort the rectangles by their origins (firstly by the Y coordinate, then the X coordinate). We tested whether this pre-processing step provides actual benefits to the model's ability to learn. We propose two more methods of ordering the input: (1) random order, (2) grouping the rectangles by their category and sorting the groups alphabetically. Figure~\ref{fig:subfig3:sorting} shows that while the alternative methods of structuring the input perform quite well on the training set, the model's ability to generalize gets diminished, which is showcased by the poor validation loss in later epochs. We conclude that since a transformer is able to extract context information based on relative positions of tokens, encoding as much information in the ordering itself should benefit the learning process.

\paragraph{Dataset Size Dependence}

The data that we use for training is procedurally generated by our data generation module. This allows us to generate as many facades as we require from the facade data distribution that the data generator represents. We use this fact to test how the amount of data our model is trained on influences its ability to reconstruct facades. We generate 4 datasets of varying sizes---starting from 20k facades all the way up to 200k facades. Since the space of all possible facade segmentations is vast, we do not expect any data-point repeats, even in the largest dataset. Our experiments show that the dataset size has a big impact on the abilities of the model---the more data it was fed, the better the results (cf. Figure~\ref{fig:subfig1:train_size}). We assume that this behavior is caused by the complexity of the function being learned which arises from subspaces, in the all possible segmentations space, that should evaluate to the same structure even though their sizings (and by consequence the sequence that represents them) differ vastly. 

\paragraph{Learning Rate and Batch Size}

We have performed multiple tests to identify what values of learning rate and batch size best fit our model. From the results (cf. Figures~\ref{fig:subfig6:lr} and \ref{fig:subfig7:batch}) we conclude that learning rates equal to 0.0001 and 0.00001 both provide satisfactory results (we opt for the smaller value), while the batch size does not impact the result substantially as long as it's not too big (we settled on a batch size of 32).

\begin{figure}[t]
        \center
        \hfill
        \begin{subfigure}[t]{0.23\textwidth}
        \centering
        \includegraphics[width=0.48\textwidth]{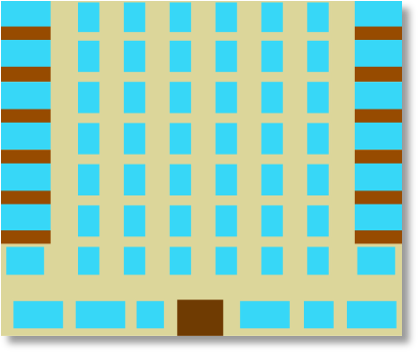}
        \includegraphics[width=0.48\textwidth]{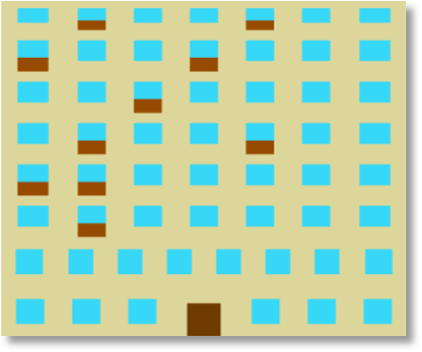}
        \caption{Failure case}
        \label{fig:sub_fail}
        \end{subfigure}
        \hfill
        \begin{subfigure}[t]{0.23\textwidth}
        \centering
        \includegraphics[width=0.48\textwidth]{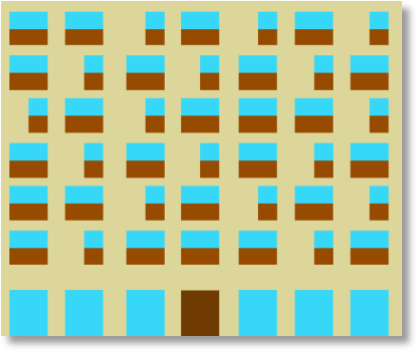}
        \includegraphics[width=0.48\textwidth]{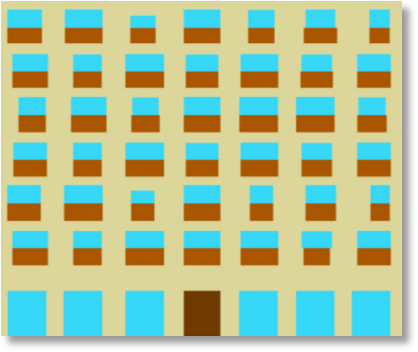}
        \caption{Limitation case}
        \label{fig:sub_limitation}
        \end{subfigure}
        \hfill
        \caption{The showcase of failure and limitation cases of our approach: (a) the model snowballs from one mistake and completely rearranges all balconies; (b) our grammar does not contain any productions which could recreate a balcony layout this irregular.}
        \label{fig:failure}
\end{figure}

\section{Discussion and Limitations}\label{sec:limit}

While our method reproduces a vast range of facades, it has limitations tied to the split-grammar language. If no derivation tree based on the grammar can reproduce a given facade, the model will not deliver a correct result (Figure~\ref{fig:failure}). This can be mitigated by extending the grammar with more procedures and making them more general, which requires insight into architectural design. Additionally, the transformer model's memory requirement scales proportionally to the sequence length squared, limiting facade sizes and detail levels.

A common failure case is the snowball effect during inference. If the model makes an error while generating the main structure, such as missing a segment on a floor, further generation can become scrambled (Figure~\ref{fig:failure}). This can sometimes be rectified by manually fixing the error and running a partial inference from that point.

\section{Conclusions and Future Work}\label{sec:conclusion}

In this paper we proposed a transformer-based model for neuro-symbolic reconstruction of procedural shape representations. We proposed a semi-complex split grammar based on previous work on procedural architecture generation~\cite{Wonka2003}. 

As a main contribution we proposed an encoder-decoder model composed of two GPT-2 transformers. We trained our model on 200,000 procedurally generated pairs of facade segmentations and split grammar based procedures. The whole dataset was automatically generated using our proposed split-grammar based generator described in Section \ref{sec:method:data_generation}. 
In our current implementation, the presented method is able to closely recreate given input facades as procedures, including their sizing parameters.

The approach could be further developed by making the grammar more general, using just a couple of multi-purpose space partition productions. This would allow the model to potentially formulate more complex functions by itself, which would be an abstraction of those primitive splits. There is also room to optimize the sequence length that is passed to the model to enable representing larger and more detailed facades. Splitting the facade into smaller parts and generating their structural representations independently is a route worth exploring. Furthermore, a natural step for the method would be reconstructing 3D buildings as procedures.

\begin{figure*}[p]
    \centering
    \input{results_figure.tex}
    \caption{Results of our method: Each sequence of 5 images first depicts the original target image and its segmentation (partially from~\cite{Musialski2012InteractiveCF}). Some of the inputs are purposefully noisy to showcase the robustness of our approach. The first image after each vertical line showcases the procedural reconstruction generated by our method, and the last two results are procedurally generated variations. Best viewed in close-up in the electronic version.}
    \label{fig:qualitative_results}
\end{figure*}

\bibliographystyle{ACM-Reference-Format}
\bibliography{bib_facade2024}

\end{document}